\documentclass[fleqn,10pt]{wlscirep}
\usepackage[utf8]{inputenc}
\usepackage[T1]{fontenc}
\usepackage{amsmath}
\usepackage{nccmath}
\title{Pleobot: a modular robotic solution for metachronal swimming}

\author[1]{Sara Oliveira Santos}
\author[1]{Nils Tack}
\author[1]{Yunxing Su}
\author[2]{Francisco Cuenca-Jiménez}
\author[2]{Oscar Morales-Lopez}
\author[2]{P. Antonio Gomez-Valdez}
\author[1,*]{Monica M. Wilhelmus}
\affil[1]{Brown University, Center for Fluid Mechanics, School of Engineering, Providence, 02912, USA}
\affil[2]{Universidad Nacional Autónoma de México, Circuito Interior s/n, Engineering, Coyoacán, 04510, México}

\affil[*]{mmwilhelmus@brown.edu}

\begin{abstract}
Metachronal locomotion is a widespread swimming mode used by aquatic swarming organisms to achieve performance and maneuverability in the intermediate Reynolds number regime. Our understanding of the mechanisms driving these abilities is limited due to the challenges of studying live organisms. Designs inspired by nature present an approach for developing small and maneuverable underwater self-propelled robots. Here, we present the design, manufacture, and validation of the \emph{Pleobot} --a unique krill-inspired robotic swimming appendage constituting the first platform to study metachronal propulsion comprehensively. Our methods combine a multi-link 3D printed mechanism with active and passive actuation of the joints to generate natural kinematics. Using force and fluid flow measurements in parallel with biological data, we show the link between the flow produced by the appendage and thrust. Further, we provide the first account of a leading-edge suction effect that contributes to lift during the power stroke. The repeatability and modularity of the \emph{Pleobot} enable the independent manipulation of particular motions and traits to test hypotheses central to understanding the relationship between form and function. Lastly, we outline future directions for the \emph{Pleobot}, including adapting morphological features. We foresee a broad appeal to a wide array of scientific disciplines, from fundamental studies in ecology, biology, and engineering, to developing new platforms for studying oceans across the solar system.
\end{abstract}
\begin{document}

\flushbottom
\maketitle
%
%
\thispagestyle{empty}

\section*{Introduction}

Nature-inspired robotics have become valuable tools in biology and ecology as they allow better-controlled comparative studies than traditional organismal methods \cite{Lauder2022}. Independent manipulation of a particular trait or behavior of interest enables testing hypotheses central to exploring the underlying mechanisms driving maneuverability and performance. While several test platforms have shed light on important aspects of biological underwater propulsion \cite{Esposito2012,Tytell2021,Howe2021,Lauder2015}, they have also produced practical engineering solutions for unmanned underwater vehicles (UUV). Previous robotic solutions for engineering problems, including tunas \cite{zhu2019tuna}, scallop \cite{Robertson2019}, and dolphins \cite{Wu2019} demonstrate improved propulsive efficiency and optimal operation in a range of environments. However, because of experimental or technical challenges, most advances have been for relatively high Reynolds numbers (Re). Consequently, despite the tremendous diversity of species inhabiting the moderate realm, we still need a comprehensive understanding of aquatic propulsion and its derived engineering applications at low-to-intermediate Re.

Metachronal swimming is a ubiquitous swimming mode among many of the most abundant invertebrate species inhabiting the transitional flow regime (Re $<$ 1000), such as copepods and krill. This propulsive gait is characterized by the sequential beating of several closely spaced swimming appendages, phase-shifted in time, in a tail-to-head traveling wave \cite{vogel2020life}. Metachrony enables large-scale behaviors in krill, particularly the ability to perform diel vertical migrations (DVM) from the sea surface to oxygen minimum zones located up to 1 km at depth \cite{bianchi_diel_2013,bianchi_global_2016}. Efficient swimming is achieved by leveraging drag during properly tuned power-and-recovery strokes \cite{vogel2020life, Kils1981}. During the power stroke, the swimming appendages (pleopods) move opposite the swimming direction while maximizing their surface area. This generates drag in each pleopod, producing thrust to propel the organism. Conversely, during the recovery stroke, the profile area decreases, thus reducing the drag on the appendages and creating a net thrust force sufficient to overcome the drag on the body. Previous works showed drag-based metachronal swimming is more effective than lift-based propulsion at intermediate Re for accelerating, braking, and turning because it generates significant thrust over short periods \cite{Byron2021,vogel2020life,vogel2003comparative,walker_mechanical_2000}.

Fluid flow experiments on live krill have established the foundations of our understanding of the effect of fluid-structure interactions on the far-field flow \cite{murphy_metachronal_2011,Murphy2013}. Murphy et al. characterized the swimming kinematics of live krill and linked the success of the propulsion system to appendage morphology, stroke kinematics, and the resulting hydrodynamic effects \cite{murphy_metachronal_2011}. Tip vortices forming on the pleopods of krill \cite{Murphy2013} and mantis shrimp \cite{Garayev2021} were shown to be central in the production of thrust \cite{kim_characteristics_2011,devoria_vortex_2012}. Kim and Gharib extended these results using idealized pleopod shapes with simplified kinematics. They demonstrated that the area over the surface of the appendage enclosed by the tip vortex, specifically, modulates thrust production \cite{kim_characteristics_2011}. However, challenges in characterizing the near-field flow and measuring the forces generated during swimming have hindered efforts to establish the relationship between fluid dynamics and force distribution (i.e., thrust and lift) in intermediate Re metachronal swimmers.

Simplified robotic models and numerical simulations have complemented these efforts, shedding light on the role of varying Re, phase lag, and appendage spacing on the hydrodynamics \cite{Alben2010, ford_hydrodynamics_2019, Garayev2021, Granzier-Nakajima2020, hayashi_metachronal_2020,lim_kinematics_2009}. For instance, Ford et al. showed that at low Re, the individual jets produced by adjacent pleopods do not interact due to viscous dissipation, but at Re around 800, they form a near-steady jet \cite{ford_hydrodynamics_2019}. The outcome of this difference is an increase in vertical and horizontal momentum that enables krill to generate thrust and lift forces required for locomotion and hovering. Phase lag also emerges as an important factor contributing to metachronal swimming efficiency, as it yields near-maximal efficiency and thrust, and achieves higher average body velocity when compared to synchronous motions \cite{Alben2010, ford_hydrodynamics_2019, hayashi_metachronal_2020, zhang_neural_2014}. Likewise, appendage spacing is a critical morphological factor linked to stroke kinematics. Previous works demonstrated that greater swimming performance is achieved when the ratio of appendage spacing to appendage length is less than one \cite{Ford2021}. This was hypothesized to be the direct consequence of inter-pleopod fluid flow interactions enhancing the vortex strength and circulation during the power stroke \cite{Garayev2021}.

While the insights gathered through simplified systems are invaluable, we still need a unifying theory of drag-based metachronal swimming to explain the relationship between appendage kinematics, fluid flows, and force production. For instance, even the production of lift needed for hovering still needs to be explored. Adding complexity to better match the natural biological and hydrodynamic phenomena is necessary to elucidate the effects of morphological and kinematic characteristics on the near- and far-field flow. We should then adopt an integrative framework allowing for the comprehensive study of the link between locomotor kinematics, form, and function.

Motivated by a robotics-inspired biology approach to address this problem, we present the first fully articulated, multi-link, metachronal robotic appendage reproducing the swimming kinematics of krill. Our novel design achieves active control of both the proximal and distal appendages and the passive out-of-plane actuation of the two rami of the distal appendages through the interaction between the pleopod and the surrounding fluid. In addition, we implemented the main morphological features of krill pleopods through the assembly of 3D-printed modules. Using force and fluid flow measurements, we quantify the effects of particular flow features on thrust and lift production and disentangle their role during the power and recovery phases of a stroke. We use biological data to validate our model as a suitable test platform. In this study, we leverage the \emph{Pleobot} to study the near-field flow of a metachronal pleopod for the first time and provide evidence for a previously unreported leading-edge suction effect that contributes to lift during the power stroke. The repeatability and modularity of the \emph{Pleobot} enables a range of configurations and motion programs to investigate metachrony across Re, taxa, and morphological attributes. As such, we anticipate the \emph{Pleobot} will broadly appeal to a wide array of scientific disciplines, such as ecology, biology, and engineering.

\section*{Results}

\subsection*{Pleobot kinematics}
\begin{figure}
    \centering
    \includegraphics[width=1\textwidth]{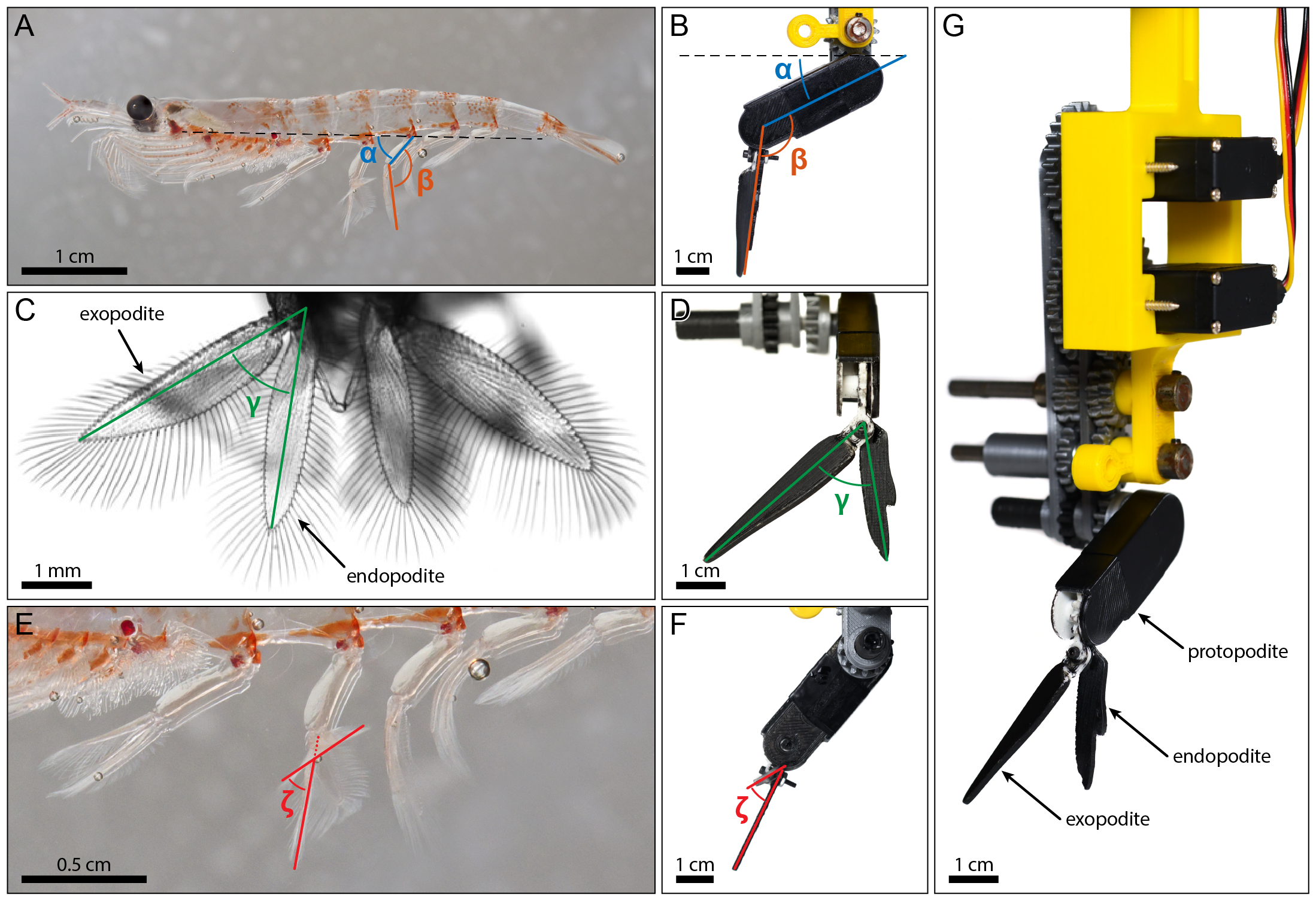}
    \caption{Morphology and kinematic parameters of the pleopod. Panels A (\emph{Euphausia superba}), C (\emph{Palaemonetes paludosus}) and E show the kinematic parameters of free-swimming shrimp incorporated in the \textit{Pleobot} presented in panel B, D and F: $\alpha$ is defined as the angle between the axis of the body and the proximal segment (protopodite), $\beta$ is the angle between the protopodite and the distal biramous segment (formed by the endopodite and the exopodite), $\gamma$ appears during the power stroke as the exopodite and endopodite separate, and $\zeta$ characterizes the cupping formed between the exopodite and the endopodite. The \textit{Pleobot} (panel G) is engineered based on a mechanical gear train to actively control $\alpha$ and $\beta$, while passively integrating $\gamma$. Note that panel C was captured from the back of the organism, while panel E shows a close up view of the pleopods from the side (as in panel A).}
    \label{fig:model}
\end{figure}

The swimming kinematics of the \emph{Pleobot} are characterized by three main angles: 1) $\alpha$, the angle between the body axis and the protopodite; 2) $\beta$, the angle between the protopodite and the biramous distal appendage (including the endopodite and exopodite); and 3) $\gamma$, the angle between the endopodite and exopodite (Fig. \ref{fig:model}A-D). The cupping angle, $\zeta$, is also implemented in the \emph{Pleobot} as seen in \emph{Euphausia superba} (Fig. \ref{fig:model}E,F). The resulting krill-inspired model (Fig. \ref{fig:model}G) uses a train of gears to actuate both $\alpha$ and $\beta$ kinematics (Fig. \ref{fig:model}B). $\gamma$ is actuated passively due to hydrodynamic forces during the power and recovery strokes (Fig. \ref{fig:model}D), and the $\zeta$ angle of the exopodite is fixed (Fig. \ref{fig:model}F).

The kinematics of the \emph{Pleobot} were validated against organismal data \cite{murphy_metachronal_2011} by tracking the motion of the robotic system over several consecutive cycles (n = 5). The cycle-averaged $\alpha$ and $\beta$ angles over time were in line with kinematics data for \emph{E. superba} (Fig. \ref{fig:Kinematics}A,B) \cite{murphy_metachronal_2011}. The slight deviation of the $\beta$ angle from krill data was associated with the tolerance between gears, which is directly proportional to the number of units in the train. The motion of the passively actuated exopodite ranged from 14$^{\circ}$ to 95$^{\circ}$, with abduction occurring at the end of the return stroke and adduction coinciding with the start of the recovery strokes (Fig. \ref{fig:model}). Comparative values over a complete beat cycle have not been reported for \emph{E. superba}, but Murphy and collaborators\cite{murphy_metachronal_2011} described $\gamma$ oscillating between 0$^{\circ}$ and 77$^{\circ}$. The kinematic analysis of the \emph{Pleobot} demonstrates it can perform the prescribed motion program with minimal error and agrees with the values reported for live krill. As such, \emph{Pleobot} is an accurate and highly repeatable robotic analog of the swimming appendages of live metachronal organisms \cite{murphy_metachronal_2011}.

\subsection*{Flow visualization}

\begin{figure}
    \centering
    \includegraphics[width=0.85\textwidth]{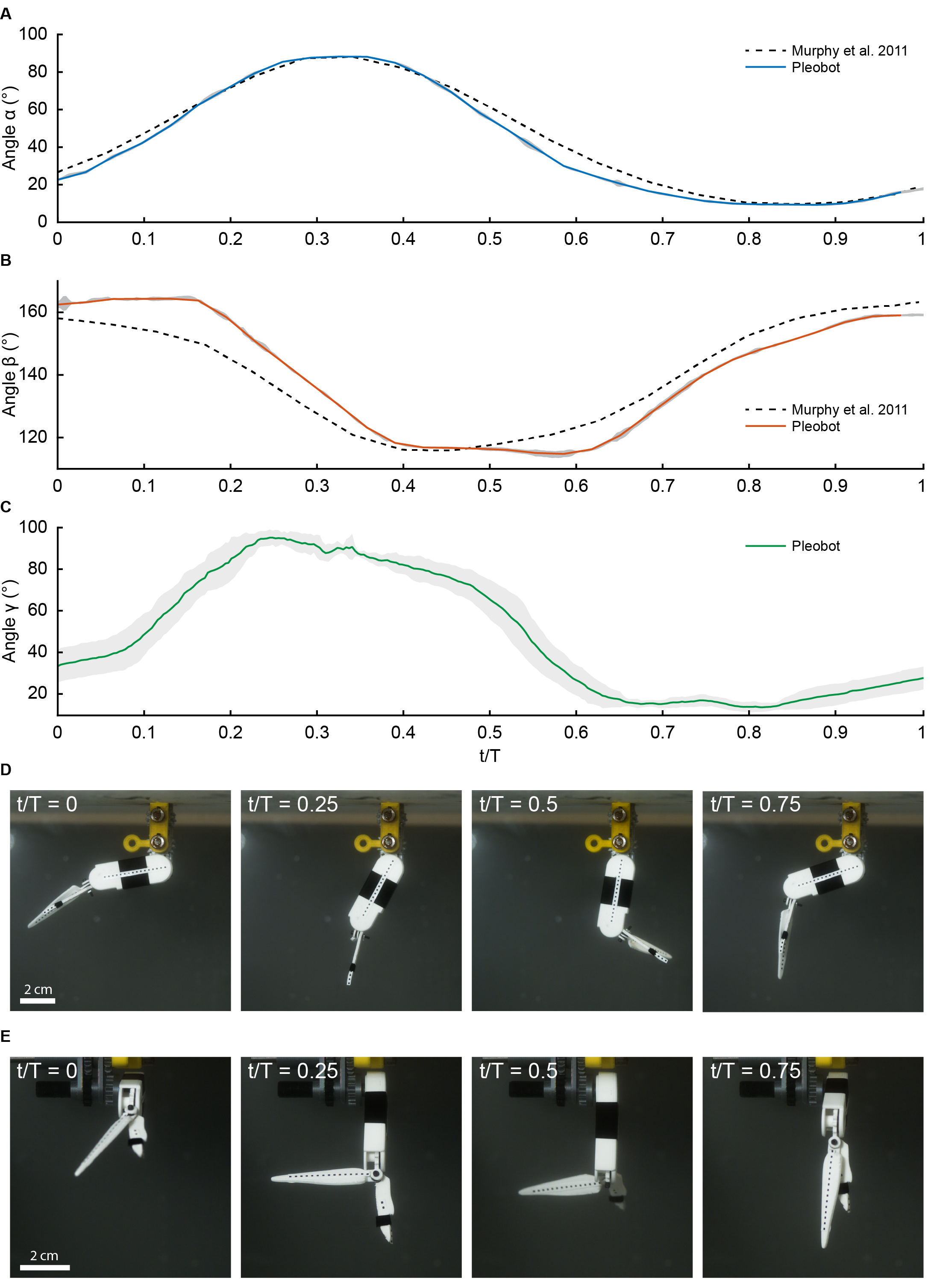}
    \caption{Kinematic measurements. Validation of the \textit{Pleobot} was achieved by tracking its motion and plotting each kinematic parameter alongside biological data \cite{murphy_metachronal_2011}. Panels A-C show the evolution of kinematic angles $\alpha$, $\beta$, and $\gamma$. Note that the standard deviation over 5 cycles is indicated as shaded regions. Panels D and E show the evolution of these angles from the lateral (panel D) and frontal (panel E) views.}
    \label{fig:Kinematics}
\end{figure}

\begin{figure}
    \centering
    \includegraphics[width=0.7\textwidth]{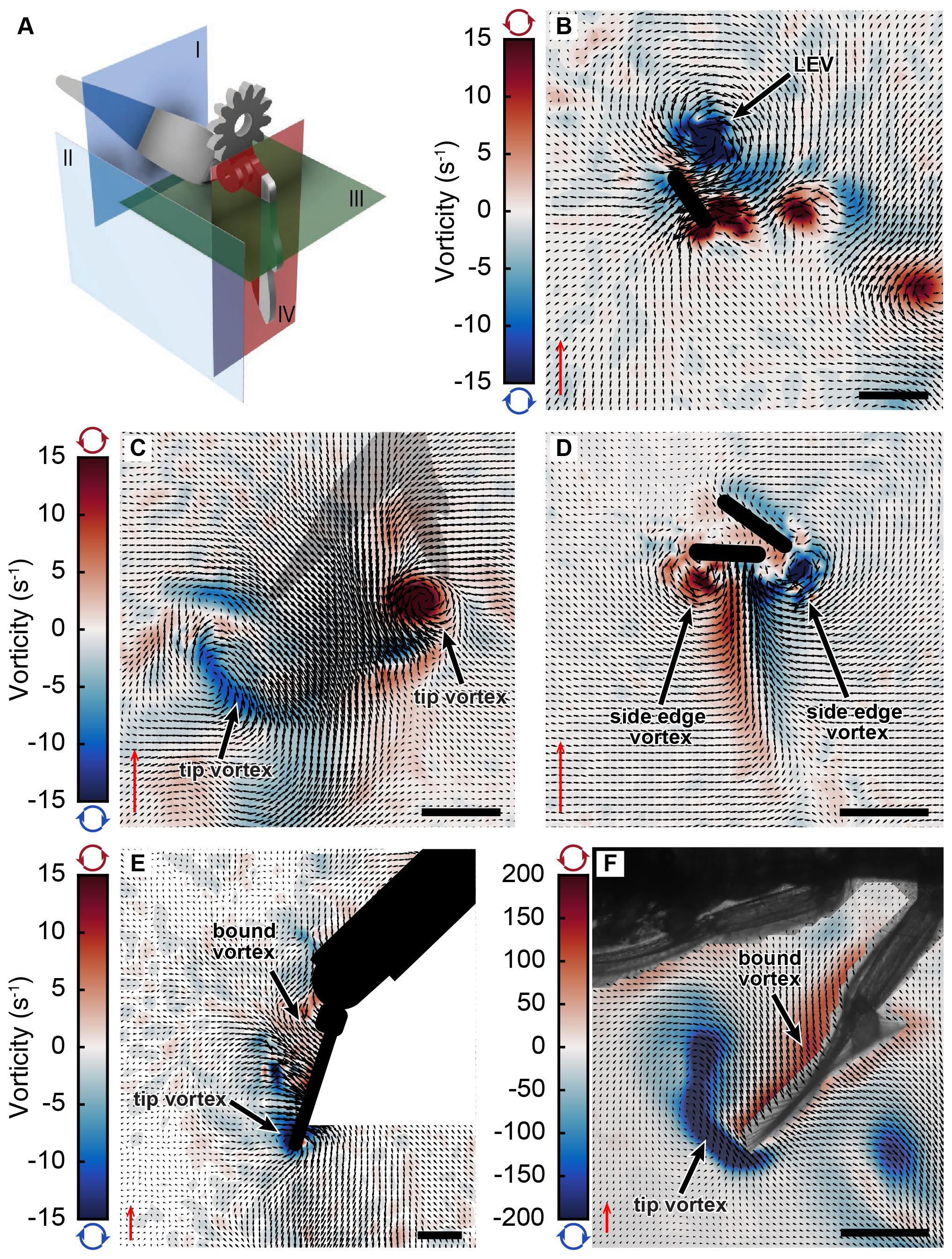}
    \caption{Flow field measurements. The planes at which PIV data were acquired are presented in panel A. Representative instantaneous velocity and vorticity fields are shown in panels: B for the vertical plane I, C for the frontal plane II, D for the horizontal plane III, and E for the vertical plane IV. Flow field measurements acquired from a tethered live shrimp are presented in panel F for reference. The black regions in panels B, and D-E are the cross-sectional profiles of the \textit{Pleobot} corresponding to the planes in A and the white area in panel E represents a shadow area of the laser. In C, the \textit{Pleobot} is out-of-plane and its position is superimposed for clarity. Note that the size and direction of the arrows scale with the value of the velocity at a given grid point, while the colors scale with the value of the vorticity according to the colormap. The red scale arrow represents 20 cm s$^{-1}$. The scale bar represents 1 cm in (B–E) and 1 mm in (F).
}
    \label{fig:PIV}
\end{figure}

PIV measurements were performed in four different planes to evaluate the flow field (Fig. \ref{fig:PIV}A). At the beginning of the power stroke (t/T, the non-dimensional stroke time, is 0), a vortex forms at the tip of the endopodite accompanied by a bound vortex uniformly distributed along most of the length of this segment (Fig. \ref{fig:PIV}E). This pattern agrees with biological data acquired using live ghost shrimp (Fig. \ref{fig:PIV}F). Meanwhile, the exopodite abducts as $\alpha$ increases and $\beta$ decreases (Fig. \ref{fig:Kinematics}). During the first half of the power stroke (before reaching t/T = 0.25), the exopodite maintains a low angle of attack (AoA, measured relative to the body axis, which is horizontal, in a clockwise direction), promoting the formation of an attached leading-edge vortex (LEV) along the anterior face (in the swimming direction) of the exopodite (Fig. \ref{fig:PIV}B). The LEV remains attached until the exopodite reaches AoA = 73$^o$ ± 1.9$^o$ at t/T = 0.3. As $\alpha$ angle increases, the AoA of the exopodite also increases to become vertical relative to the swimming direction (AoA = 90$^o$) during the second half of the power stroke (Fig. \ref{fig:Force}I). Given that the pleopod decelerates during this phase, the LEV separates from the exopodite (Figs \ref{fig:PIV}C, \ref{fig:Force}I) and the bound vortex of the endopodite sheds at the tip (Figs \ref{fig:PIV}E,F, \ref{fig:Force}E). The combined effect of the flow entrained by the two counter-rotating vortices results in the production of a large posteriorly oriented downward jet.

Upon initiating the recovery stroke, the exopodite adducts and overlaps with the endopodite to effectively reduce the overall profile area of the pleopod subject to the flow (Figs \ref{fig:Kinematics}C,E, \ref{fig:PIV}D, \ref{fig:Force}C). Contrary to the power stroke, this configuration generates a pair of counter-rotating side edge vortices that induces spanwise flow in the swimming direction (Figs \ref{fig:PIV}D, \ref{fig:Force}F,G).

\subsection*{Hydrodynamic forces}

\begin{figure}
    \centering
    \includegraphics[width=1\textwidth]{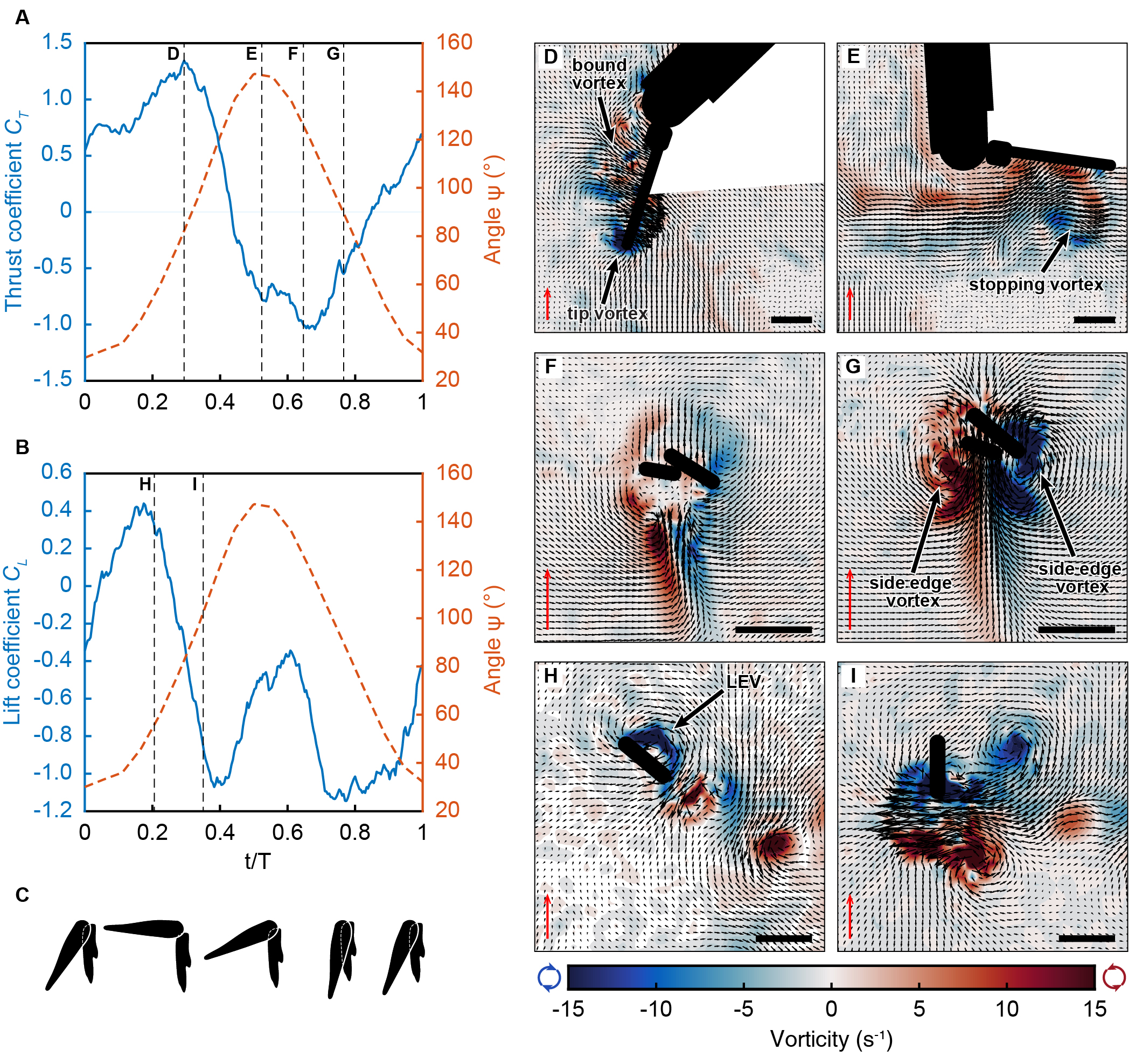}
    \caption{Force measurements. The thrust (A) and lift (B) coefficients are averaged over five consecutive cycles. The orange dashed lines show the kinematics of the appendage, characterized by the angle $\Psi$. Corresponding pleopod profiles in panel C for the anterior view of the $\gamma$ angle emphasize the change in pleopod surface area over a beat cycle . Panels D-I show PIV measurements at different times of a cycle, corresponding to the black dashed lines in panels A and B. Panels D and E correspond to the lateral view along the endopodite during the power stroke, showing a tip and bound vortex (D) that are shed at the end of the power stroke (E). Panels F and G show a bottom view of the exopodite and endopodite during the recovery stroke, with strengthening spanwise flow behind the appendages between two side edge vortices. Panels H and I correspond to a lateral view along the midplane of the exopodite during the power stroke, where we identify a leading-edge vortex (LEV) that initially contributes primarily to lift and then thrust as the AoA changes. The scale bar and the red scale arrow represent 1 cm and 20 cm s$^{-1}$, respectively for all frames.
 }
    \label{fig:Force}
\end{figure}

Force measurements were performed to evaluate the contribution of lift and thrust forces during steady forward swimming (Fig. \ref{fig:Force}). The lift ($C_L$) and thrust ($C_T$) coefficients were computed for a 20x-scaled endopodite-and-exopodite model. At the beginning of the power stroke (t/T = 0), there is a significant amount of thrust force, likely because of the incoming spanwise flow produced behind the pleopod at the end of the previous beat (Fig. \ref{fig:Force}G). During the power stroke, thrust and lift strongly correlate with the abduction of the exopodite. The forces on the endopodite decompose in the thrust and lift directions, and the LEV forming along the exopodite drives this association. Throughout the first third of the power stroke (t/T $\sim$ 0.16), the lift coefficient increases sharply to reach its global maximum ($C_L$ = 0.44) as the exopodite is abducting and produces an attached LEV (Fig. \ref{fig:Force}B,C,H). In contrast, while thrust is being produced, it remains mostly constant during this phase, and only increases significantly after lift peaks and decreases (Fig. \ref{fig:Force}A). This inverse relationship coincides with the AoA of the exopodite shifting to a more vertical orientation that causes the LEV to shed. It also results from the increased contribution of the endopodite to thrust, given the overall vertical orientation of the entire pleopod (see Fig. \ref{fig:Force}A,B). During the second half of the power stroke, thrust decreases sharply as the bound vortex of the endopodite moves toward the tip and forms a stopping vortex (Fig. \ref{fig:Force}E). Given the increase of the effective angle of the pleopod relative to the flow (angle $\Psi$, the overall angle between the body axis and the frontal face of the endopodite; see Fig. \ref{fig:Force}), the endopodite and the exopodite produce a downward force inducing minimum lift. The subsequent increase in lift force is likely due to the occurrence of spanwise flow entrained by the stopping vortex that pushes against the anterior face of the decelerating endopodite and exopodite (Fig. \ref{fig:Force}E). 

During the first 2/3 of the recovery stroke, thrust is negative, suggesting this phase of a beat is mostly dominated by drag forces. The thrust coefficient is lowest around the mid-recovery stroke (Fig. \ref{fig:Force}A) when the pleopod accelerates. Thrust increases gradually to become positive, thus indicating some thrust is produced even during the recovery stroke. After the start of the recovery stroke, lift drops sharply to another minimum with decreasing $\psi$ angle (Fig. \ref{fig:Force}B). Lift rises again at the end of the recovery stroke when the appendage decelerates. Increasing thrust and lift coincide with the occurrence of spanwise flow behind the overlapping endopodite and exopodite (Fig. \ref{fig:Force}G).

\section*{Discussion}

In recent years, the growing need for maneuverable AUVs for underwater exploration has galvanized approaches inspired by nature \cite{aubin2019, katzschmann2018,neveln2013,wang2019, white2021}. In particular, the discovery of extraterrestrial oceans motivates the development of novel robotic platforms that will likely require the efficiency, versatility, and maneuverability of metachronal swimming. As such, the \emph{Pleobot} lays down the foundations for the upcoming work on metachronal swimming to inform the design of underwater explorers.

\emph{Pleobot} represents the first fully articulated robot to study metachronal swimming. The design constitutes a modular, 3D-printed platform incorporating the kinematic and morphological characteristics of krill -- a marine organism observed to form large aggregations and migrate hundreds of meters into the ocean. While krill kinematics have been explored in great detail, a quantitative analysis of the fluid flow near individual appendages is needed to link kinematics to hydrodynamics and understand force production. \emph{Pleobot} is the first system to disentangle the roles of kinematic parameters, morphology, and near-field hydrodynamics on thrust and lift generation at the scale of a single appendage.

Our unique study shows that \emph{Pleobot} serves as an analog system to investigate a wide range of parameters and characteristics that collectively contribute to the efficiency and flow characteristics of metachronal propulsion. Although krill has been used as a model organism in biological, computational, and robotic studies, the focus has mostly been on the flow field of several appendages rather than the singular mechanisms for the generation of thrust and lift. Conclusions regarding fluid-structure interactions are thus solely possible by measuring the near field flow and by introducing complexities in the model that enable independent testing of hypotheses relevant to the ecology and behavior of the natural systems (i.e., maneuverability, migrations, fast swimming). 

In our study, the analysis of force and PIV data allows us to evaluate the contribution of specific flow features to the generation of lift and thrust in \emph{Pleobot}. At the beginning of the power stroke, the endopodite contributes to thrust and lift through the formation of tip and bound vortices. Similarly, the exopodite also contributes to the generation of both forces. Creation of lift during steady forward swimming is necessary for krill because they are negatively buoyant and must create lift to maintain their position in the water column \cite{Murphy2013}. As the stroke progresses, the exopodite becomes more vertically oriented and its contribution to lift decreases, while its contribution to thrust increases (Fig. \ref{fig:Force}A,B). This represents a trade-off between the generation of lift and thrust production. While the primary purpose of thrust in steady swimming is to propel the organism forward, lift is needed to guarantee the vertical position in the water column. Our data suggest that rather than being determined by behavioral processes, thrust and lift production are constrained to a relatively narrow range of kinematics and morphological criteria. Behavioral changes in the gait and beat frequency invariably modulate these forces. However, the dependency on the $\psi$ angle of the passive abduction of the exopodite and its orientation in the flow demonstrates that a limited set of system requirements are fundamentally important for thrust and lift production in metachronal propulsion.

Notably, we reveal the existence of a LEV along the exopodite during the power stroke. Force measurements show that lift is enhanced by the formation of this attached LEV on the exopodite during the power stroke. Furthermore, this vortex delays the loss in lift forces slightly to maintain positive lift throughout part of the power stroke. This lift-based mechanism likely supplements the effects of the drag-based mechanism stemming from the pleopod pushing on water. These results in the near field illustrate the importance of the \emph{Pleobot}, not only as a platform to engineer solutions for ocean exploration in the intermediate Reynolds number regime but also as a framework to understand nature. Force measurements on a biologically-inspired appendage highlight the importance of understanding the vortex dynamics around the pleopod and reveal that both lift and thrust are generated simultaneously.

\begin{figure}
    \centering
    \includegraphics[width=0.8\textwidth]{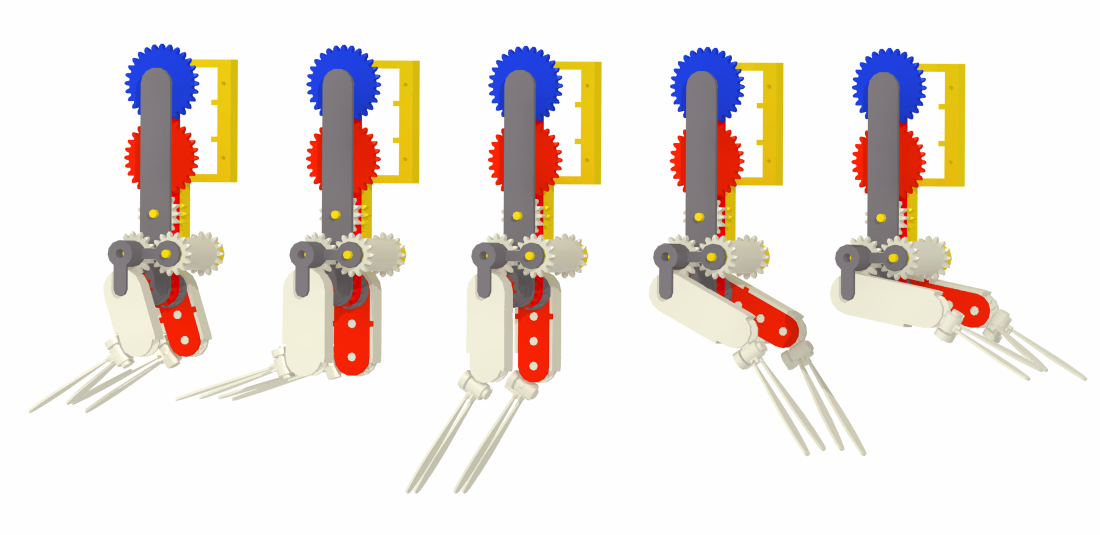}
    \caption{Scaled KRill Inspired Modular Pleobots (SKRIMP). This sketch illustrates how the modularity and rapid prototyping of the \emph{Pleobot} enables the investigation of metachronal swimming by employing several units. This includes fluid flows and forces - in several taxa, body plans, and appendage configurations. \emph{SKRIMP} constitutes the baseline to engineer a new generation of AUVs operating in complex marine environments leveraging the swimming characteristics of metachronal swimmers. }
    \label{fig:RkCAD}
\end{figure}

The incorporation of natural kinematics and morphological traits into the \emph{Pleobot} provides promising results and incentives for increasing the complexity of our robotic model and elucidate additional mechanisms contributing to efficiency in krill and other metachronal swimmers. Markedly, the presence of hair-like setae around the endopodite and exopodite, and flexibility of the propulsors have the most potential for altering the flow field around the pleopods. Setae are defining features in metachronal organisms in the low-to-intermediate Re that increase the surface area in contact with water to enhance thrust \cite{Koehl2004, vogel2020life, cheer_paddles_1987}. Flexibility has been shown to be an important characteristic of propulsors in smoothing out thrust peaks to generate nearly constant thrust during the power stroke \cite{kim_characteristics_2011}. \emph{Pleobot} provides a platform to evaluate the importance of these additional morphological features. Moreover, its modular design can be easily replicated and grouped in several units to investigate mechanisms such as constructive pleopod interactions (Fig. \ref{fig:RkCAD}). For example, the proximity of several appendages beating at a phase lag promotes fluid flow interactions that likely contribute to enhanced thrust \cite{Ford2021b, Garayev2021}. The \textit{Pleobot} emerges as an ideal platform to quantify these effects.

Finally, while organismal investigations form the core of our knowledge about metachronal swimming \cite{Byron2021,Garayev2021,murphy_metachronal_2011,Murphy2013}, working with live animals provides only a partial understanding of the biomechanical and hydrodynamic mechanisms leading to efficiency and performance. Recent mechanical and robotics approaches allowing repeatable measurements highlight the valuable contribution of low-cost, controllable test platforms for biological studies \cite{Ford2021,ford_hydrodynamics_2019,Ford2021b}. Such a comparative approach will be central to developing a unifying theory of metachronal swimming. Manufacturing of the \emph{Pleobot} is accessible to both institutions and individuals who want to experiment (see \verb|oliveira:zenodo:2022)|). Its modular design allows for the quick prototyping of different appendage shapes and sizes. Furthermore, the control of both its proximal and distal segments facilitates the study of a wide range of systems, from small copepods, to krill, and larger species like mantis shrimp and lobsters. The \emph{Pleobot} can thus be employed to facilitate the comprehensive characterization of metachronal, drag-based propulsion and help establish a unifying theory for this locomotor mechanism. 

\section*{Methods}

\subsection*{Modeling of krill appendage kinematics}

\noindent
To reproduce the kinematics of \emph{Euphausia superba}, as reported by Murphy et al.\cite{murphy_metachronal_2011}, we use values for $\alpha$ and $\beta$ to find kinematic relationships for \textit{Pleobot}. The \textit{Pleobot} is dynamically scaled based on the Reynolds number. This dimensionless parameter is defined using the velocity and the length scale of the appendages and is approximately 600 for live krill:

\begin{ceqn}
\begin{align}
 Re = \frac{U_{tip}*L}{\nu} = \frac{2 \theta n L^2}{\nu} \;
\end{align}
\end{ceqn}

\noindent
where $L$ is the length of the pleopod, $U_{tip}$ the average velocity at its tip, and $\nu$ the kinematic viscosity of the fluid. $U_{tip}$ is calculated using the stroke amplitude $\theta$, the frequency $n$, and the length $L$ as $2\theta n L$. 

The locomotor system is implemented by a multi-link mechanism using a transmission gear box actuating both the proximal and distal segments (corresponding to $\alpha$ and $\beta$) with independently controlled servo motors. Angle $\gamma$ represents out-of-plane motion, posing challenges for active actuation and is thus passively actuated via the hydrodynamic interaction between the structure (pleopods) and the fluid.

The kinematic relationships governing the positions of the gears are\cite{erdman1997mechanism,sandor1984advanced}:%
\begin{ceqn}
\begin{align}
\frac{\Delta \phi _{k+1}-\Delta \phi _{k}}{\Delta \phi _{k-1}-\Delta \phi
_{k}}=-\frac{N_{k-1}}{N_{k+1}}=-\dfrac{r_{k-1}}{r_{k+1}} \;
\end{align}
\end{ceqn}
\noindent
This equation gives the relationship between the driving and driven gears, $\Delta \phi _{k+1}$ and $\Delta \phi _{k-1}$, respectively, and the link connecting both, $ \Delta \phi _{k}$ (Supplementary figure 1), where $N$ and $r$ are the number of teeth and radius of the corresponding gear. The position of the pleopod, $\mathbf{R}_{pi}$,  is governed for a given appendage $i$ as (Supplementary figure 2,B):%
\begin{ceqn}
\begin{align}
\mathbf{R}_{pi} = \mathbf{R}_{1i}+\mathbf{R}_{2i} \;
\end{align}
\end{ceqn}

\noindent
where%
\begin{ceqn}
\begin{align}
\begin{tabular}{lll}
$\mathbf{R}_{1i}=\mathbf{R}\left( \theta _{1i}\right) \mathbf{r}_{1i}$ ;&   
$\mathbf{R}_{2i}=\mathbf{R}\left( \theta _{2i}\right) \mathbf{r}_{2i}$ \\  
$\mathbf{r}_{1i}=\left[ x_{1i},0\right] ^{T}$ ; &
$\mathbf{r}_{2i}=\left[ x_{2i},0\right] ^{T}$ \\ 
\end{tabular}
\end{align}
\end{ceqn}

The local reference frame vectors for the protopodite and exopodite are $\mathbf{r}_{1i}$ and $\mathbf{r}_{2i}$. Similarly, $\mathbf{R(\theta_{1i})}$ and $\mathbf{R(\theta_{2i})}$ are the global rotation matrices and $x_{1i}$ and $x_{2i}$ the lengths (Supplementary figure 2,A).

Finally, the angular displacements of the gears are obtained using the relationships described in Equation 2:%
\begin{equation}
\frac{\Delta \psi _{2i}-\Delta \theta _{1i}}{\Delta \psi _{1i}-\Delta
\theta _{1i}} = -r_{e1} ,\\
\frac{\Delta \psi _{3i}-\Delta \theta _{1i}}{\Delta \psi _{2i}-\Delta
\theta _{1i}} = -r_{e2} ,\\ 
\frac{\Delta \psi _{4i}-\Delta \theta _{1i}}{\Delta \psi _{3i}-\Delta
\theta _{1i}} = -r_{e3} 
\end{equation}%
\noindent
where $\psi$ is the rotation of the gear along link 1 (endopodite), $\theta$ is the angle of the first link measured from the global reference frame (Supplementary figure 2,A), and $r_e$ is the gear ratio. Equations (3) and (5) allow for the solution of direct kinematics of the mechanism by using $\alpha _{i}~$and $\beta _{i}$, to calculate $\psi _{1i}$, which moves the second bar (endopodite)(Supplementary figure 2,C). The first bar
(protopodite) is moved by angle $\alpha _{1i}$ (Supplementary figure 2,A). The metachronal trajectory at the end of pleopod 1, is given by $\mathbf{R}_{p1}=\left[
x_{p1},y_{p1}\right] ^{T}$.

Manufacturing was completed by 3D printing, keeping small tolerances to reduce vibrations and loss of movement of the gear train. The supports house the servos and gears, and were designed to be an above-water structure. Bearings are used to reduce friction between gears and the axes. The transmission has an amplification  of 2.5 to achieve the desired angular speed for the links. 

\noindent 
\subsection*{Exopodite and endopodite}
As \emph{E. superba} propels forward, the endopodite and exopodite abduct and adduct to effectively change the profile area of the appendages to generate thrust and reduce drag. This motion is characterized by $\gamma$, the angle between the exopodite and the endopodite. In the \textit{Pleobot}, this process is actuated on the horizontal plane, lateral to the protopodite.

In forward-swimming krill, exopodite abduction occurs at the beginning of the power stroke, and reaches a maximum of 77 $^{\circ}$ \cite{murphy_metachronal_2011}. This motion induces cupping of the appendages creating a V-shaped structure \cite{Murphy2013}, reminiscent of those observed in swimming fish that have been shown to produce greater thrust compared to flat fins (e.g., see \cite{Esposito2012}). The exopodite and endopodite adduct during the recovery stroke.

Appendage cupping forms angle $\zeta$ between the endopodite and the exopodite. Photographic evidence was used to quantify $\zeta$ by measuring the angle between the midplanes of the endopodite and the exopodite (Fig. \ref{fig:model}E) . The \textit{Pleobot} was set to match the mean quantified value of 37$^{\circ}$.

\noindent
\subsection*{Robotic design}
Modular CAD designs were printed with polylactic acid (PLA) using a Prusa i3 MKS3+ 3D printer (Prusa Research, Prague, Czech Republic) for fast prototyping. Each pair of appendages is actuated by two servos (HS-5087MH, Hitec RCD, San Diego, CA, USA), controlled by a microcontroller (ELEGOO Mega 2560, Elegoo Industries, Shenzhen, China) programmed using Simulink (MathWorks, Natick, MA, USA) via two repeating sequence interpolated blocks, one for $\alpha$ and one for $\beta$, prescribing the angles adapted from \cite{murphy_metachronal_2011}.  The full CAD library and assembly as well as the list of purchased components can be accessed in the open-access repository by Oliveira Santos et al. (see \verb|oliveira:zenodo:2022)|.

The swimming kinematics of the \textit{Pleobot} were analyzed and compared to those reported for live krill \cite{murphy_metachronal_2011}. One robotic appendage was tethered to a traverse beam and submerged in a glycerin-water mixture at room temperature (60$\%$ glycerin and 40$\%$ water). Appendage motion was recorded at 125 fps using a scientific camera (FASTCAM MINI WX, Photron, 2048 pixels x 2048 pixels). Black markers on the surface of the robotic pleopod were tracked, both on the protopodite and the exopodite by digitizing the video recordings via DLTdv8 for MATLAB using automatic point tracking \cite{Hedrick_2008}.

\noindent
\subsection*{Flow field measurements}
The flow field around the beating appendage was measured using 2D PIV. The experiments were carried out in a cubic tank of 30 cm by side using a high-speed camera (FASTCAM MINI WX, Photron, 2048 pixels x 2048 pixels) at 125 frames per second, a Nikon lens (Nikon AF-S VR Micro-NIKKOR 50 mm), and a continuous laser (Laserglow, 1 W at 532 nm) with a cylindrical lens to create a laser sheet. The flow was seeded with 10 $\mu$m particles (Dantec Dynamics, Skovlunde, Denmark) (Supplementary Figure 3). The field of view is 174 mm x 174 mm, resulting in a spatial resolution of 0.085 mm per pixel. The velocity fields were calculated using DaVis 10 (LaVision) with decreasing interrogation window sizes (initial 48 pixels x 48 pixels, 50$\%$ overlap, 1 pass, and final 32 pixels x 32 pixels, 50$\%$, 3 passes). Standard vector post-processing was performed to remove outliers in the flow field. The vorticity field was calculated from the velocity field.

\noindent
\subsection*{Force measurements}
The force measurements were conducted on a scaled-up (20x) model of the distal appendage by using a 6-axis force transducer (Nano 17 F/T transducer, ATI) set up in a cubic tank of 30 cm by side (Supplementary Figure 4). The Reynolds number was matched to ensure  dynamically scaled experiments. To this end, the beating frequency and the fluid viscosity were modified from the baseline. A glycerin-water mixture (60$\%$ glycerin and 40$\%$ water) was used with a dynamic viscosity of 9 cSt, measured with a standard rheometer (Ares-G2, TA Instruments). The appendage was designed to be neutrally buoyant in the water-glycerin mixture.

The force transducer was mounted in the axis of rotation of the bi-ramous pleopod, such that the x-axis was aligned with the pleopod, and the y-axis was perpendicular to it. The forces measured, parallel ($F_\parallel$) and perpendicular ($F_\bot$) to the pleopod, are decomposed into lift and thrust by taking into account the orientation of this section in the body via the  angle $\Psi$, the overall angle between the body axis and the endopodite, defined as $\alpha + 180^o - \beta$. The lift and thrust coefficients ($C_L$ and $C_T$, respectively) are calculated using the lift and thrust forces measured from the decomposed components (Equation 6). Using the instantaneous force along a cycle, averaged over five cycles, we calculate lift and thrust as:

\begin{ceqn}
\begin{align}
\begin{tabular}{c}
$Lift = F_\parallel \ sin \ \psi - F_\bot \ cos \ \psi$  \\
$Thrust = F_\parallel \ cos \ \psi + F_\bot \ sin \ \psi$
\end{tabular}
\end{align}
\end{ceqn}
and $C_L$ and $C_T$ as

\begin{ceqn}
\begin{align}
\begin{tabular}{c c}
    $C_L = \frac{Lift}{\frac{1}{2} \rho U^2 A }$ & ; $C_T =  \frac{Thrust}{\frac{1}{2} \rho U^2 A}$
\end{tabular}
\end{align}
\end{ceqn}

where $\rho$ is the density of the glycerin-water mixture, $U$ is the appendage tip velocity, and A the area of the abducted appendage.
\\

\noindent
\subsection*{Organism experiments}
Live ghost shrimp \emph{Palaemonetes paludosus} were used to validate the hydrodynamics and $\gamma$ angle kinematics of the \emph{Pleobot}. Ghost shrimp are not obligate swimmers like krill, but they are appropriate analogs for several reasons. First, they have a ubiquitous reliance on metachronal swimming. Second, they have comparable morphology and swimming kinematics (i.e., parameters related to their exopodites, endopodites, and setae) to krill. Finally, they are similar in body size and anatomy.

Specimens (n = 4) were tethered to a vertical wire using cyanoacrylate applied to the dorsal surface of their carapace and hung in a filming vessel (15 × 10 × 5 cm$^3$) brought to 100\% oxygen saturation. We performed bright-field PIV by seeding the water with 10 $\mu$m particles (Dantec Dynamics, Skovlunde, Denmark) and back-lighting with a fiber optic illuminator (Fiber-Lite MI-152, Dolan-Jenner Industries, Boxborough, MA, USA) coupled with a collimating lens (N-BK7 Plano-Convex Lens, Thorlabs, Newton, NJ, USA) \cite{Gemmell2014}. We recorded the lateral and posterior views using a high-speed digital video camera (Fastcam Nova R2, Photron, Tokyo, Japan) at 2000 fps and a resolution of 2048 × 1472 squared pixels.

Velocity vectors were calculated using the DaVis 10 software package (LaVision, Göttingen, Germany). Image pairs were analyzed with three passes of overlapping interrogation windows (75\%) with decreasing size (96 × 96 to 64 × 64 squared pixels). All frames were used for analysis, yielding a time separation between frames of 0.5 ms. Masking of the body and pleopods before image interrogation confirmed the absence of surface artifacts in the PIV measurements.

\bibliography{references,scibib}
\section*{Acknowledgements}

We gratefully acknowledge Valentina Di Santo for insightful discussions that strengthened the foundation of this work. We also thank Yair Sanchez Juarez and Pedro Enrique Ávila Hernández for their work on the preliminary versions of the \textit{Pleobot}. Funding for this research was provided by the University of California Institute for Mexico and the United States (UC MEXUS), the Consejo Nacional de Ciencia y Tecnologia (CONACYT) through the UC MEXUS-CONACYT collaborative grants program (UCMEXUS CN-18-138), Brown University OVPR Research Seed Grant, and NASA Rhode Island EPSCoR Seed Grant  (80NSSC22M0040).

\section*{Author contributions statement}
 M.M.W., S.O.S, F.C.J, P.A.J.V. and O.M.L developed the concept and designed experiments; S.O.S., N.T, Y.S. performed experiments and analyzed data; M.M.W., S.O.S, F.C.J, P.A.J.V., O.M.L., N.T and Y.S. wrote the manuscript; M.M.W. supervised the project. M.M.W. administered the project. All authors reviewed the manuscript. 

\section*{Additional information}

\textbf{Competing interests} The authors declare no competing interests.\\
\textbf{Data Availability Statement} Data will be provided upon reasonable request to the corresponding author via email.

\end{document}


\renewcommand{\figurename}{Supplementary figure}
\begin{figure}
    \centering
    \includegraphics[width=0.6\textwidth]{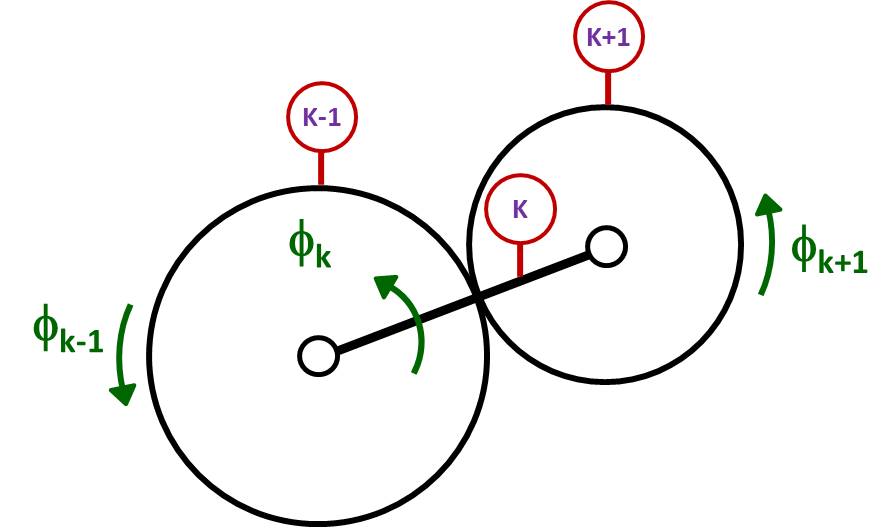}
    \caption{Epicyclic gear pair. Gear pair with input rotation $\phi$ and link number $k$. We obtain the relationship between the driving and driven gears, $\Delta \phi _{k+1}$ and $\Delta \phi _{k-1}$, respectively, and the link connecting both, $ \Delta \phi _{k}$. Image adapted from \cite{sandor1970kinematic}
 }
 \end{figure}

 \begin{figure}
    \centering
    \includegraphics[width=1\textwidth]{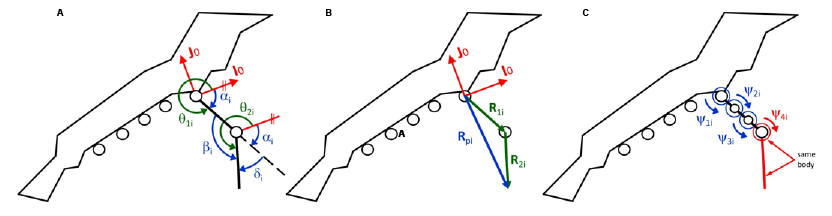}
    \caption{The locomotive system of the \emph{Pleobot}. Panel A shows a diagram that includes the global reference frame of one pleopod, $J_o$ and $I_o$, along with angles $\alpha_i$, the angle between the body axis and the proximal appendage as measured in krill, $\beta_i$, the angle between the proximal and distal segment as measured in krill, $\theta _{1i}=2\pi -\alpha _{i}$,  $\theta _{2i}=\theta _{1i}-\delta_{i}=\pi +\beta _{i}-\alpha _{i}$, and $\delta _{i}=\pi -\beta _{i}$. Panel B shows the position of the pleopod, $\mathbf{R}_{pi}$, which is $\mathbf{R}_{pi} = \mathbf{R}_{1i}+\mathbf{R}_{2i}$, where $\mathbf{R}_{1i}=\mathbf{R}\left( \theta _{1i}\right) \mathbf{r}_{1i}$ and $\mathbf{R}_{2i}=\mathbf{R}\left( \theta _{2i}\right) \mathbf{r}_{2i}$. Here $\mathbf{r}_{1i}$ and $\mathbf{r}_{2i}$ represent the local reference frame vectors, and $\mathbf{R(\theta_{1i})}$ and $\mathbf{R(\theta_{2i})}$ are the global rotation matrices for the protopodite and endopodite, respectively. Panel C shows angle $\Psi$, the rotation of the gear along link 1 (endopodite), where $\Delta \theta _{1i}=\theta _{1i}-\theta _{1i,0}$, $\Delta \psi _{1i}=\psi _{1i}-\psi_{1i,0}$, $\Delta \psi _{2i}=\psi _{2i}-\psi _{2i,0}$, $\Delta \psi_{3i}=\psi _{3i}-\psi _{3i,0}$, $\Delta \psi _{4i}=\Delta \theta_{2i}$, and $\Delta \theta_{2i}=\theta _{2i}-\theta _{2i,0}$.
 }
 \end{figure}

 \begin{figure}
    \centering
    \includegraphics[width=1\textwidth]{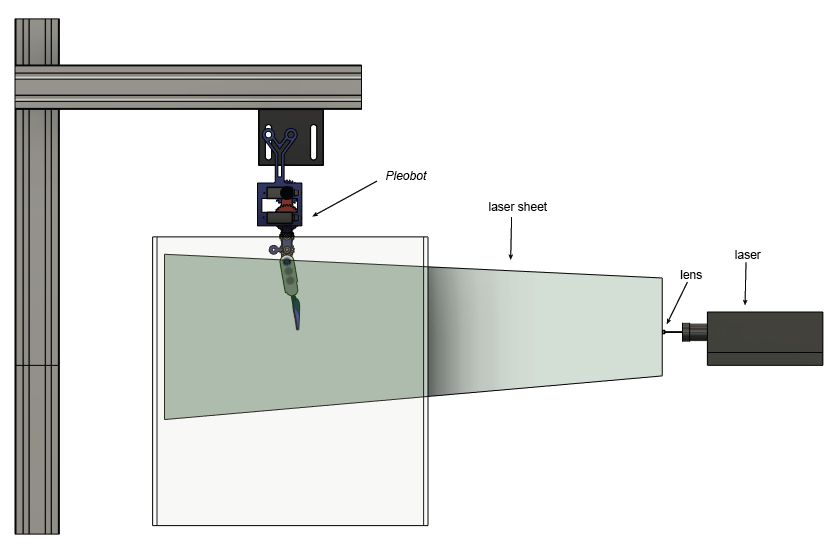}
   \caption{The flow field around the beating appendage was measured using 2D PIV. The experiments were carried out in a cubic tank of 30 cm by side using a high-speed camera (FASTCAM MINI WX, Photron, 2048 pixels x 2048 pixels) at 125 frames per second, a Nikon lens (Nikon AF-S VR Micro-NIKKOR 50 mm), and a continuous laser (Laserglow, 1 W at 532 nm) with a cylindrical lens to create a laser sheet. The flow was seeded with 10 $\mu$m particles (Dantec Dynamics, Skovlunde, Denmark).}
 \end{figure}

 \begin{figure}
    \centering
    \includegraphics[width=1\textwidth]{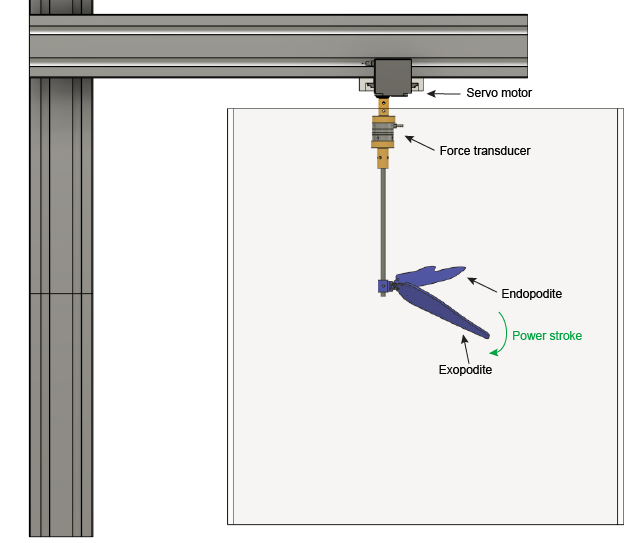}
    \caption{Experimental setup for force measurements. The force measurements were conducted using a scaled-up (20x) model of the distal appendage (endopodite and exopodite). This was mounted onto a 6-axis force transducer (Nano 17 F/T transducer, ATI) set up in a cubic tank of 30 cm by side. The kinematics were prescribed by the angle $\Psi$ calculated as $\alpha + 180^o - \beta$. The Re (1500) was used to dynamically scale the experiments by decreasing the beating frequency and increasing the fluid viscosity. A glycerin-water mixture (60$\%$ glycerin and 40$\%$ water) was used with a dynamic viscosity of 9 cSt, measured with a standard rheometer (Ares-G2, TA Instruments). The appendage was designed to be neutrally buoyant in the water-glycerin mixture.}
 \end{figure}

\bibliography{scibib}

\bibliographystyle{Science}